\documentclass{kluwer}    
\input{epsf.tex}

\newdisplay{guess}{Conjecture}

\begin{document}                                                              
                   
\begin{article}
\begin{opening}    
\title{Chaos in cuspy triaxial galaxies with a supermassive black hole:
a simple toy model}
\author{Henry E. \surname{Kandrup}\email{kandrup@astro.ufl.edu}}
\institute{Department of Astronomy, Department of Physics, and Institute
for Fundamental Theory, University of Florida, Gainesville, 
Florida, USA 32611}
\author{Ioannis V. \surname{Sideris}\email{sideris@astro.ufl.edu}}
\institute{Department of Astronomy, University of Florida, Gainesville, 
Florida, USA 32611}
\runningauthor{H. E. Kandrup and I. V. Sideris}
\runningtitle{Chaos in cuspy triaxial galaxies}

\date{\today}

\begin{abstract}
This paper summarises an investigation of 
chaos in a toy potential which mimics much of the 
behaviour observed for the more realistic triaxial generalisations of the 
Dehnen potentials, which have been used to model cuspy triaxial galaxies both 
with and without a supermassive black hole. The potential is the 
sum of an anisotropic harmonic oscillator potential, 
$V_{o}={1\over 2}(a^{2}x^{2}+b^{2}y^{2}+c^{2}z^{2})$, 
and a spherical Plummer potential,
$V_{p}=-M_{BH}/\sqrt{r^{2}+{\epsilon}^{2}}$, with $r^{2}=x^{2}+y^{2}+z^{2}$.
Attention focuses on three issues related to the properties of ensembles of
chaotic orbits which impact on chaotic mixing and the possibility of
constructing self-consistent equilibria: (1) What fraction of the orbits are
chaotic? (2) How sensitive are the chaotic orbits, {\em i.e.}, how large are 
their largest (short time) Lyapunov exponents? (3) To what extent is the 
motion of chaotic orbits impeded by Arnold webs,  
{\em i.e.}, how `sticky' are the chaotic orbits? These questions are explored
as functions of the axis ratio $a:b:c$, black hole mass $M_{BH}$, softening 
length ${\epsilon}$, and energy $E$ with the aims of understanding how the 
manifestations of chaos depend on the shape of the system and why the black
hole generates chaos. The simplicity of the model makes it amenable to a
perturbative analysis. That it mimics the behaviour of more complicated 
potentials suggests that much of this behaviour should be generic.
\end{abstract}
\keywords{galaxies: kinematics and dynamics}

\end{opening}

\section{Motivation}
Over the past decade it has become apparent that many early-type galaxies
are genuinely three-dimensional objects, neither spherical nor axisymmetric 
({\em cf.} \citeauthor{KaB} \citeyear{KaB})
that they typically have a central density cusp 
({\em cf.} \citeauthor{lau} \citeyear{lau})
and that the center of the galaxy often contains a supermassive black hole 
({\em cf.} \citeauthor{KaR} \citeyear{KaR}).
However, there is substantial numerical evidence 
({\em cf.} \citeauthor{FaM} \citeyear{FaM})
that the combination of triaxiality 
and a cusp typically yields potentials that admit a large amount of chaos; 
and there is also evidence indicating that the introduction of a supermassive 
black hole into a cuspy, triaxial potential causes a further increase in the 
abundance of chaos 
({\em cf.} \citeauthor{VaM} \citeyear{VaM}).
Numerical explorations indicate further 
({\em cf.} \citeauthor{SaK} \citeyear{SaK})
that the phase space 
associated with a cuspy triaxial potential, either with or without a black 
hole, can be very complex, laced with an intricate Arnold \shortcite{arn}
web that can seriously impede phase space transport, so that chaotic
orbits in these potentials can be very `sticky' in the sense recognised
originally by Contopoulos \shortcite{con}. 

However, there are at least two important questions which have not yet been
addressed adequately. Why does the combination of triaxiality and a
cusp and/or black hole lead to chaos? And to what extent is the observed
behaviour generic? Almost all work hitherto has focused on one class
of model potentials, namely the triaxial generalisations of the spherical
Dehnen \shortcite{deh} potentials first considered by 
\citeauthor{MaF} \shortcite{MaF}.

\citeauthor{MaV2} \shortcite{MaV1} have argued that a black hole or cusp tends 
to induce chaos because it serves as a near-singular perturbation that 
destabilises what would be regular box orbits if the galaxy had a smooth core 
and there were no black hole. This seems quite reasonable. However, 
\citeauthor{udr} \shortcite{udr}
have noted a fact well known from nonlinear dynamics 
({\em cf.} \citeauthor{lal} \citeyear{lal}), namely
that chaos often arises simply by breaking a symmetry or combining incompatible
symmetries. One might therefore argue that a supermassive black hole acts as 
a source of chaos because of symmetry-breaking: the unperturbed galaxy is
triaxial, whereas the perturbation manifests spherical symmetry. This could,
{\em e.g.}, account for the fact that, as a source of chaos, even a 
comparatively low mass black hole can be more important than a steep cusp. As 
\citeauthor{VaM} \shortcite{VaM} noted, `even a modest black hole, with a mass 
${\sim}{\;}0.3\%$ the mass of the galaxy, is as effective 
as the steepest central density cusp at inducing stochastic diffusion.'
In any event, one knows that a black hole or cusp does not always result in 
chaos. By carefully maintaining symmetries, one can, {\em e.g.}, construct 
integrable generalisations of the Staeckel potentials which incorporate a
supermassive black hole \cite{st1,st2}.

One objective here is to address the question of why cusps and black holes 
induce chaos by studying orbits in an extremely simple class of potentials 
which appear to exhibit behaviour qualitatively similar to what is observed 
the triaxial Dehnen potentials with or without a black hole. The potential,
\begin{equation}
V(x,y,z)={1\over 2}(a^{2}x^{2}+b^{2}y^{2}+c^{2}z^{2})-
{M_{BH}\over \sqrt{r^{2}+{\epsilon}^{2}}},
\end{equation}
with $r^{2}=x^{2}+y^{2}+z^{2}$, is the sum of an anisotropic harmonic 
potential and a spherical Plummer potential. There are five parameters which 
can be varied independently, namely $a$, $b$, and $c$, which 
determine the frequencies and the shape of the unperturbed oscillator, 
$M_{BH}$, the mass of the black hole, and the softening length ${\epsilon}$. 
That much of the behaviour exhibited by orbits in the triaxial 
Dehnen potentials can be reproduced by such a simple model suggests that it 
may be generic. That the model is so simple implies that it is amenable to an 
analytic treatment with the black hole viewed as a perturbation of the 
integrable oscillator potential. 

The numerical component of the analysis focuses on three important 
characterisations of the chaos: (1) What fraction of the orbits are chaotic?
(2) How large is the largest Lyapunov exponent associated with the chaotic
orbits? (3) How `sticky' are the chaotic orbits? Can they travel relatively
unimpeded throughout the accessible phase space, or is their motion obstructed
significantly by the Arnold web? 

The relative abundance of chaotic orbits is important given the general 
presumption that one requires a substantial number of regular, or nearly
regular, orbits to serve as a skeleton for the equilibrium ({\em cf}. 
\citeauthor{bin} \citeyear{bin}). 
The degree of sensitivity exhibited by orbits is important because it impacts 
on the efficacy of chaotic mixing 
({\em cf}. \citeauthor{KaM} \citeyear{KaM}, \citeauthor{MaV1} 
\citeyear{MaV2}).
The existence of ``sticky'' orbits is of potential importance 
because these orbits can in principal be used to support regular structures 
in phase space regions where few, if any, regular orbits exist 
({\em cf.} \citeauthor{woz} \citeyear{woz}, \citeauthor{kac} \citeyear{kac},
\citeauthor{paq} \citeyear{paq}).

Determining the answers to these questions as functions of the axis ratio
$a:b:c$ and the black hole mass $M_{BH}$ provides information about how
the manifestations of chaos vary with the degree of triaxiality and the
mass of the black hole. Determining the effects of varying ${\epsilon}$ 
can shed insights into why the chaos arises. For sufficiently small 
softening, the exact value of ${\epsilon}$ is immaterial, but when 
${\epsilon}$ becomes too large its value begins to have an important
effect.

If the only function of the black hole were to act as a near-singular
perturbation, one might expect that, for ensembles of orbits with fixed
energy $E$, unperturbed frequencies $a$, $b$, and $c$, and softening 
${\epsilon}$, the relative abundance of chaotic orbits should exhibit a very
simple dependence on $M_{BH}$. For $M_{BH}=0$ and $M_{BH}\to\infty$ the
potential is integrable. However, for other values $V$ is nonintegrable and,
for intermediate values far from $M_{BH}=0$ and ${\infty}$, one might expect
large measures of chaotic orbits. In particular, one might expect that a
plot of $n(M_{BH})$, the relative measure of chaotic orbits, as a function
of $M_{BH}$ would exhibit a single maximum at an intermediate value where
$M_{BH}$ is neither too small nor too large. This is {\em not} what the
numerical simulations reveal. Rather, one finds that the abundance of chaos
varies in a more irregular fashion which, as discussed in Section 2, would
suggest that chaos arises from a resonance overlap between the natural
frequencies of the oscillator and the black hole.


Section 2 of this paper focuses on the statistical properties of short time 
Lyapunov exponents. Section 3 summarises an analysis that focused primarily 
on the Fourier spectra of both individual orbits and orbit ensembles. Section 
4 concludes by summarising the results of Sections 2 and 3 and commenting on 
their physical implications.

\section{Lyapunov Exponents for Orbit Ensembles}
\subsection{WHAT WAS COMPUTED}
Most of the experiments focused on the statistical properties of ensembles of 
orbits evolved for comparatively short times. However, these were 
supplemented by very long time integrations
of individual initial conditions used to extract estimates of the
values of the largest Lyapunov exponents for fixed values of frequencies $a$, 
$b$, $c$, mass $M_{BH}$, softening ${\epsilon}$, and energy $E$. Three 
different questions were addressed:
\par\noindent
1. For fixed $a$, $b$, $c$, $M_{BH}$, and $E$, how do things depend on 
${\epsilon}$? For sufficiently small ${\epsilon}$, the precise value is 
immaterial but for larger ${\epsilon}$ this is no longer true. 
\par\noindent
2. For fixed $a$, $b$, $c$, ${\epsilon}$, and $E$, how do things depend on 
$M_{BH}$? 
The most naive picture might suggest that $n(M_{BH})$, the relative measure 
of chaotic orbits, should vary smoothly with $M_{BH}$ and that a plot of $n$ 
as a function of $M_{BH}$ should exhibit no structure. 
\par\noindent
3. For fixed ${\epsilon}$, $M_{BH}$, and $E$, how do things depend on the axis
ratio $a:b:c$? Conventional wisdom would suggest that larger deviations from
spherical symmetry imply more chaos, so that one might expect the relative
abundance of chaos to increase monotonically with increasing asphericity. 

The experiments were performed for frequencies $a$, $b$, and 
$c{\;}{\sim}{\;}1$ and energies $E{\;}{\sim}{\;}0.1-1$, this corresponding
physically to a galaxy of mass $M{\;}{\sim}{\;}1$, linear size 
$r{\;}{\sim}{\;}1$, and characteristical orbital, or dynamical, time scale 
$t_{D}{\;}{\sim}{\;}2-3$. The black hole mass was taken 
in the range $10^{-4}{\;}{\le}{\;}M_{BH}{\;}{\le}{\;}1$. Smaller black hole 
masses are not large enough to induce significant stochasticity; much larger
masses seem completely unrealistic. 

For parameters in these general ranges, the exact value of ${\epsilon}$ seems 
unimportant, provided only that ${\epsilon}^{2}<10^{-3}$ or so. For this 
reason, most of the experiments were performed assuming 
${\epsilon}^{2}=10^{-4}$, the same value used in Siopis \& Kandrup (2000). A 
`typical' choice of frequencies was $a^2=1.25$, $b^2=1.0$, and $c^2=0.75$, 
corresponding to a moderately triaxial configuration.

For fixed frequencies $a$, $b$ and  $c$, mass $M_{BH}$, and energy $E$, the 
value of the largest Lyapunov exponent  was estimated by selecting six or more
initial conditions corresponding to chaotic orbits and integrating each initial
condition for a total time $t=256000$ while simultaneously tracking the 
evolution of a small initial perturbation periodically renormalised in the
usual way ({\em cf.} \citeauthor{lal} \citeyear{lal}).
The relative abundance of
chaotic orbits was estimated by sampling the phase space to generate a 
collection of $2000$ initial conditions and evolving each of these initial
conditions, along with a small perturbation, for a time $t=2048$ which, for 
the energies considered here, corresponded to ${\sim}{\;}800t_{D}$. The value
of the largest short time Lyapunov exponent ${\chi}$ 
({\em cf.} Grassberger {\em et al} 1988)
was then used as a diagnostic to determine whether or not 
the initial condition was chaotic. (Integrations for shorter times would seem 
better motivated astronomically, but it was found that, for much smaller $t$, 
it was often impossible to determine with any reliability whether or not any 
given segment corresponded to a chaotic orbit. Indeed, even for integrations
for times as long as $t=2048$ it is possible that some very sticky chaotic
segments were misinterpreted as regular segments!) The initial conditions were 
generated as in \citeauthor{SaK} \shortcite{SaK}, using a scheme which 
generalises Schwarzschild's \shortcite{sch} algorithm to construct a 
representative library of orbits. 

The resulting short time ${\chi}$'s were examined to determine three things, 
namely (1) the relative number of chaotic orbits; (2) the degree to which 
different chaotic orbit segments with the same energy do, or do not, have 
the same short time ${\chi}$'s, and (3) the degree to which all the chaotic
segments are characterised by short time ${\chi}$'s well separated from zero.
The behaviour of orbit segments over shorter, more astronomically relevant,
time scales was addressed by partitioning each of the $2000$ segments of 
length $t=2048$ into eight segments of length $t=256$ and extracting a short
time Lyapunov exponent for each segment 
({\em cf.} \citeauthor{KaM} \citeyear{KaM}). The
resulting $16000$ segments were analysed to determine a distribution of short
time Lyapunov exponents, $N[{\chi}(t=256)]$, which probes the varying degrees 
of chaos exhibited by a representative collection of orbits, both regular and 
chaotic, over an interval of time comparable to the age of the Universe.
Finally, as another probe of the total amount of chaos, independent of 
distinctions amongst regular, sticky, and wildly chaotic orbits, the output
was analysed to extract the average short time Lyapunov exponent for {\em all}
the orbits in each ensemble. 

Most of the experiments focused on three representative energies, 
$E=1.0$, $0.6$, and $0.4$. A variety of axis ratios were considered. One
set of experiments involved considering frequencies $a$, $b$, and $c$ for
which $a^2:b^2:c^2=1+{\Delta}:1:1-{\Delta}$ and exploring the effects of 
increasing ${\Delta}$.
\subsection{WHAT WAS FOUND}
The black hole acts as an important source of chaos whenever conditions are
such that the force that it exerts on a test particle often becomes comparable 
to the force associated with the oscillator. For small 
${\epsilon}$, a mass $M_{BH}>M_{min}{\;}{\sim}{\;}10^{-3}$ appears required to 
induce an appreciable amount of obvious chaos. The precise value of 
${\epsilon}$ seems unimportant provided only that ${\epsilon}$ is very small 
compared with the characteristic size of a stellar orbit. If the black hole 
mass be frozen at some value $M_{BH}$ significantly above $M_{min}$ and 
${\epsilon}$ increased systematically, chaos remains common until the maximum 
black hole force 
${\sim}{\;}M_{BH}/{\epsilon}^{2}$ becomes smaller than a characteristic value 
somewhat less than unity, this corresponding to the typical force exerted
by the oscillator. It thus appears that it is not the singular character 
of the perturbation {\em per se} that is responsible for triggering chaos; 
rather what is important is that the force associated with the black hole 
can become comparable to, or larger than, the force associated with the
unperturbed oscillator. In other words, mixing two integrable potentials of
comparable strength is responsible for the chaos. 

\begin{figure}[t]
\centerline{
        \epsfxsize=8cm
        \epsffile{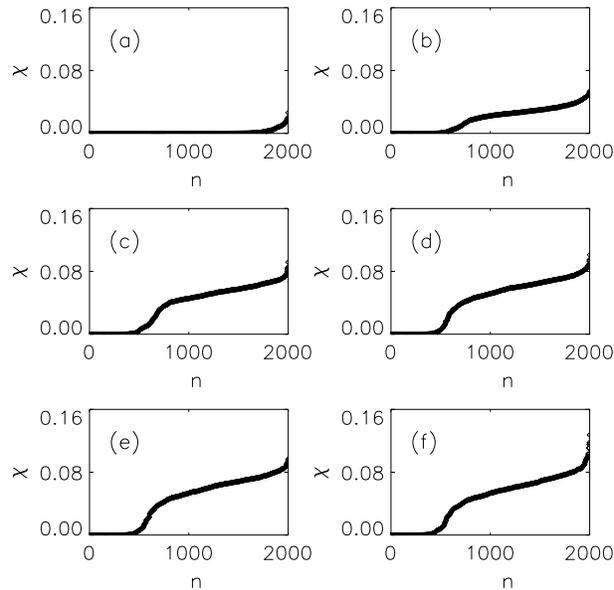}
           }
        \begin{minipage}{10cm}
        \end{minipage}
        \vskip -0.3in\hskip -0.0in
\caption
{Short time ${\chi}(t=2048)$ for $2000$ orbits evolved in the potential (1)
with $a^{2}=1.25$, $b^{2}=1.0$, $c^{2}=0.75$, $E=1.0$, and $M_{BH}=0.1$ with 
variable ${\epsilon}$.
(a) ${\epsilon}=10^{-0.5}$. (b) ${\epsilon}=10^{-1.0}$. 
(c) ${\epsilon}=10^{-1.5}$.
(d) ${\epsilon}=10^{-2.0}$. (e) ${\epsilon}=10^{-2.5}$. 
(f) ${\epsilon}=10^{-3.0}$.}
\label{FIG1}
\end{figure}


One example of how the amount of chaos depends on ${\epsilon}$ is provided by
Figure~\ref{FIG1}, 
each panel of which exhibits the computed short time ${\chi}(t=2048)$ 
for a collection of $2000$ initial conditions evolved in the potential (1) 
with 
$a^{2}=1.25$, $b^{2}=1.0$, $c^{2}=0.75$, and $M_{BH}=0.1$ with fixed energy 
$E=1.0$. Each panel was constructed by computing $2000$ short time ${\chi}$'s, 
ordering the orbits in terms of the computed ${\chi}$, and then 
({\em cf.} \citeauthor{SaK} \citeyear{SaK})
plotting the resulting ${\chi}$'s as a function of 
particle number from $1$ to $2000$. The six panels exhibit data for values of 
${\epsilon}$ ranging from ${\epsilon}^{2}=10^{-1}$ to $10^{-6}$. The computed 
${\chi}$'s for ${\epsilon}^{2}=10^{-3}$, $10^{-4}$, and $10^{-5}$ are 
virtually identical. For ${\epsilon}^{2}=10^{-2}$ chaos is weaker in two 
senses, 
namely that the relative number of chaotic orbits is smaller {\em and} that 
the typical ${\chi}$ associated with a chaotic orbit has also decreased. For 
${\epsilon}^{2}=10^{-1}$, chaos is almost completely suppressed. For 
${\epsilon}^{2}=10^{-6}$, there are a few orbits with somewhat larger 
${\chi}$'s than was observed for ${\epsilon}^{2}=10^{-5}$, $10^{-4}$, and 
$10^{-3}$, each corresponding to a particle which passed extremely close to 
the black hole. 

\begin{figure}
\centerline{
        \epsfxsize=8cm
        \epsffile{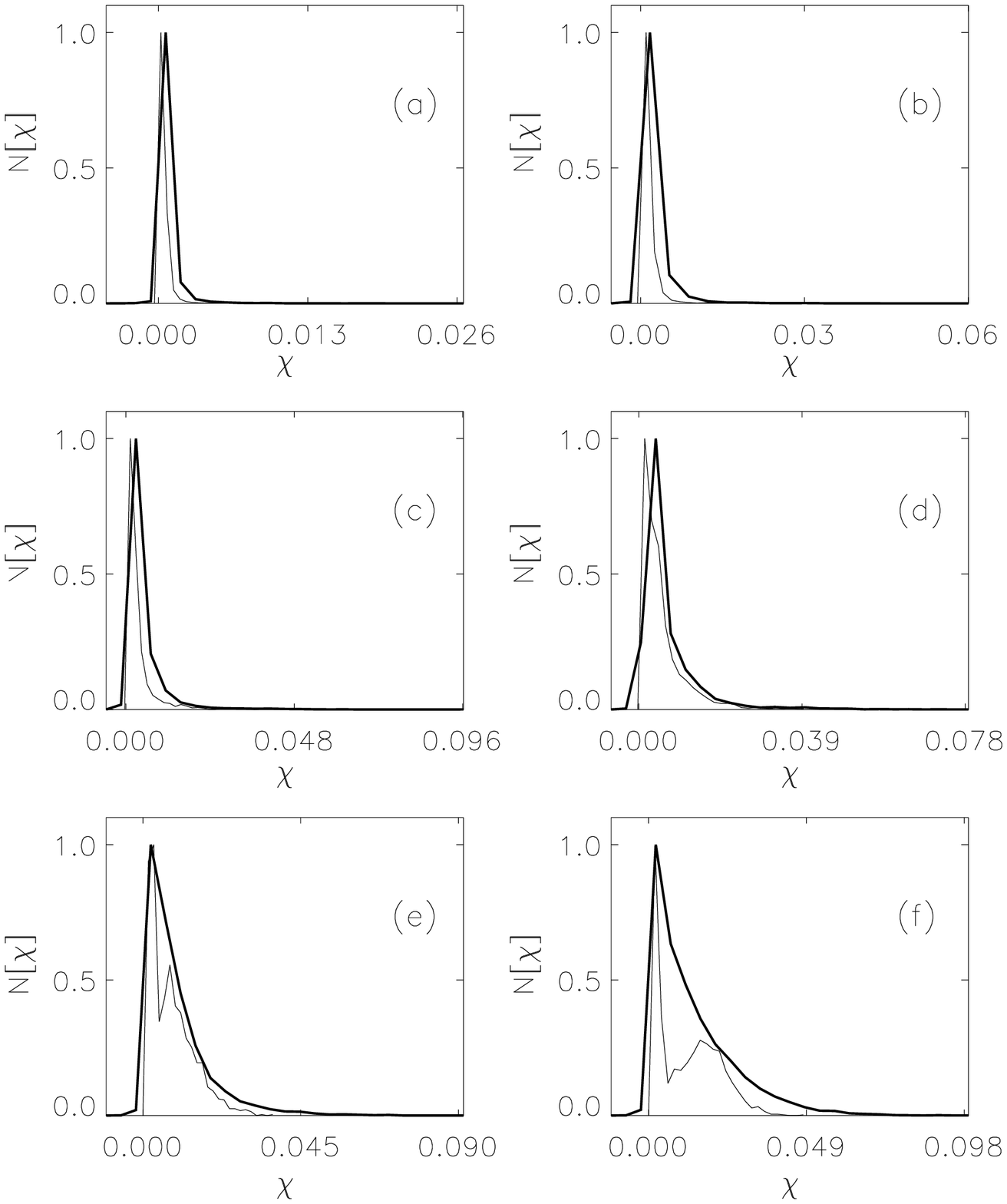}
           }
        \begin{minipage}{10cm}
        \end{minipage}
        \vskip -0.3in\hskip -0.0in
\caption{ 
Distribution of short time Lyapunov exponents, $N[{\chi}(t=256)]$ (thick
lines) and $N[{\chi}(t=2048)]$ (thin lines) generated from $2000$ 
orbit segments evolved in the potential (1) with $a^{2}=1.0$, $b^{2}=1.25$, 
$c^{2}=0.75$, $E=1.0$, ${\epsilon}=10^{-2}$, and variable $M_{BH}$. 
(a) $\log_{10} M_{BH}=-3.5$. (b) $\log_{10} M_{BH}=-3.0$. 
(c) $\log_{10} M_{BH}=-2.5$ .(d) $\log_{10} M_{BH}=-2.25$. 
(e) $\log_{10} M_{BH}=-2.0$.  (f) $\log_{10} M_{BH}=-1.75$. 
}
\label{FIG2}
\end{figure}

\begin{figure}
\centerline{
        \epsfxsize=8cm
        \epsffile{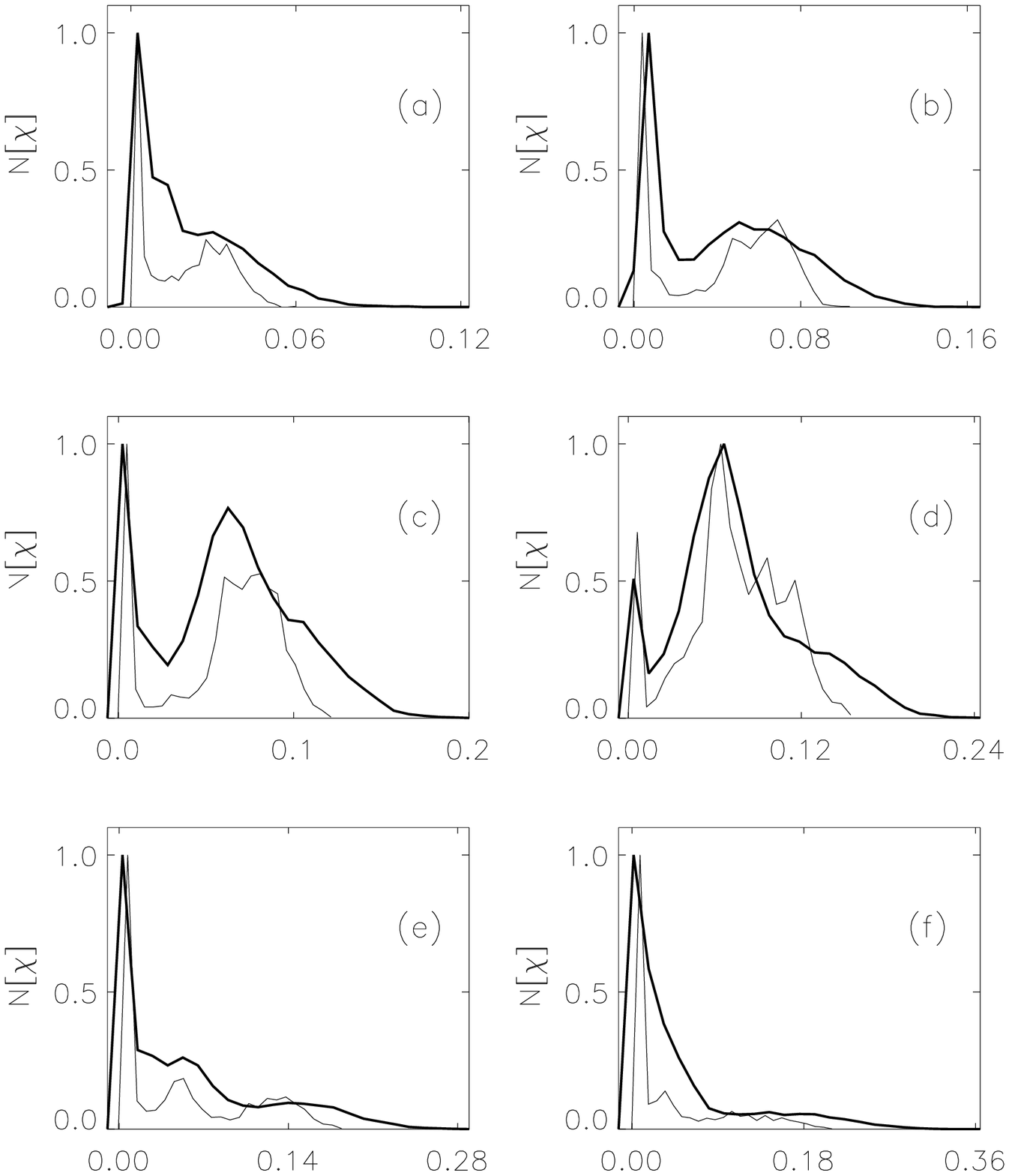}
           }
        \begin{minipage}{10cm}
        \end{minipage}
        \vskip -0.3in\hskip -0.0in
\caption{ 
Distribution of short time Lyapunov exponents, $N[{\chi}(t=256)]$ (thick
lines) and $N[{\chi}(t=2048)]$ (thin lines) generated from $2000$ 
orbit segments evolved in the potential (1) with $a^{2}=1.0$, $b^{2}=1.25$, 
$c^{2}=0.75$, $E=1.0$, ${\epsilon}=10^{-2}$, and variable $M_{BH}$. 
(a) $\log_{10} M_{BH}=-1.5$. (b) $\log_{10} M_{BH}=-1.0$. 
(c) $\log_{10} M_{BH}=-0.75$. (d) $\log_{10} M_{BH}=-0.5$. 
(e) $\log_{10} M_{BH}=-0.25$. (f) $\log_{10} M_{BH}=0$. 
}
\label{FIG3}
\end{figure}
Overall, chaotic orbits tend to be extremely sticky. In particular, even for
$t=800t_{D}$ or larger, the computed values of ${\chi}$ need not manifest a
clear and unambiguous separation between regular and chaotic behaviour. If,
as in Figure 1, an ordered list of ${\chi}$'s is plotted as a function of 
particle number, there is in general no clean break between the regular and
chaotic orbits. Oftentimes the only way to infer a reasonable demarcation 
between regular and chaotic behaviour is to look for a `kink' in the slope.
However, the relative abundance of regular/nearly regular as opposed to 
wildly chaotic orbits {\em can} be extracted by examining distributions
of short time Lyapunov exponents, $N[{\chi}(t)]$. Typical examples of such 
distributions are provided in Figures~\ref{FIG2} and~\ref{FIG3}, which exhibit 
$N[{\chi}(t)]$ for
$t=256$ and $t=2048$ for the $2000$ orbit ensembles evolved with $a^{2}=1.25$, 
$b^{2}=1.0$, $c^{2}=0.75$, ${\epsilon}= 10^{-2}$, and $E=1.0$ with variable 
black hole masses ranging between $M_{BH}=10^{-3.5}$ and $M_{BH}=1$. 

In each of these panels, the data were so plotted that the horizontal axis 
terminates at the right at a value ${\chi}_{max}$ corresponding to the largest 
value of ${\chi}(t=256)$ assumed by any of the $16000$ segments. It is thus 
evident that, for small $M_{BH}$, there are a very few orbits with 
comparatively large values of ${\chi}$, even though the overwhelming majority 
of the orbits have much smaller values. For relatively large values of 
$M_{BH}$, {\em e.g.}, $M_{BH}=10^{-0.25}$, 
there are comparable numbers of both regular/nearly regular and wildly chaotic 
orbits. For somewhat lower values, the abundance of wildly chaotic orbits 
tends to decrease with increasing $M_{BH}$, although the trend is not 
completely uniform. Thus, {\em e.g.}, there are more wildly chaotic orbits for 
$M_{BH}=10^{-0.5}$ and $M_{BH}=10^{-1.5}$ than for $M_{BH}=10^{-1}$.

For small values of $M_{BH}$, it becomes exceedingly difficult to estimate the
fraction of the orbits which are truly chaotic. In particular, for 
$M_{BH}=10^{-2}$ and less, the distribution $N[{\chi}]$ manifests a single 
sharp peak at a value close to zero. This might suggest that the orbits are 
all regular; but $N[{\chi}]$ also has a tail extending to much larger values 
which is clearly indicative of chaos. One possible interpretation is that,
{\em for small $M_{BH}$, many/all of the orbits are in fact chaotic, but that 
most of these chaotic orbits behave in a near-regular fashion almost all of 
the time}. For $M_{BH}=0$,
each initial 
condition corresponds to a regular box orbit which densely fills a region in 
configuration space that includes the origin. One might perhaps expect that,
for sufficiently small $M_{BH}$, the orbits remain ``nearly boxy'' and again
follow trajectories which occasionally bring them arbitrarily close to the
origin. In the limit ${\epsilon}\to 0$, one would thus anticipate that every
orbit eventually experiences perturbations of arbitrarily large amplitude 
which trigger chaotic behaviour. Making ${\epsilon}$ finite introduces a
bound to the perturbing force but, for ${\epsilon}$ as small as $10^{-2}$,
the value used here, this still corresponds to a maximum force of amplitude 
$10^{4}M_{BH}$. 

\begin{figure}
\centerline{
        \epsfxsize=8cm
        \epsffile{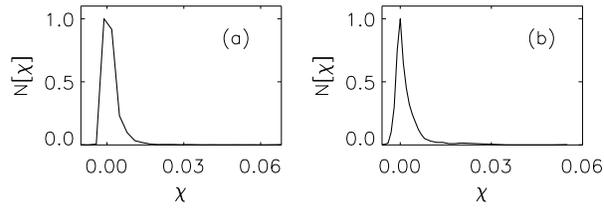}
           }
        \begin{minipage}{10cm}
        \end{minipage}
        \vskip -2.3in\hskip -0.0in
\caption{ 
(a) Distribution of short time Lyapunov exponents, $N[{\chi}(t=256)]$ 
generated from a single initial condition with $a^{2}=1.0$, $b^{2}=1.25$, 
$c^{2}=0.75$, ${\epsilon}=10^{-2}$, $E=1.0$, and $M_{BH}=10^{-3}$ integrated 
for a total time $t=8000\times 256$. The initial condition corresponded to
an orbit which for early times was wildly chaotic. (b) $N[{\chi}(t=256)]$ for 
another initial condition with the same parameter values, in this case 
corresponding at early times to a `sticky,' near-regular orbit.}
\label{FIG4}
\end{figure}

\begin{figure}
\centerline{
        \epsfxsize=8cm
        \epsffile{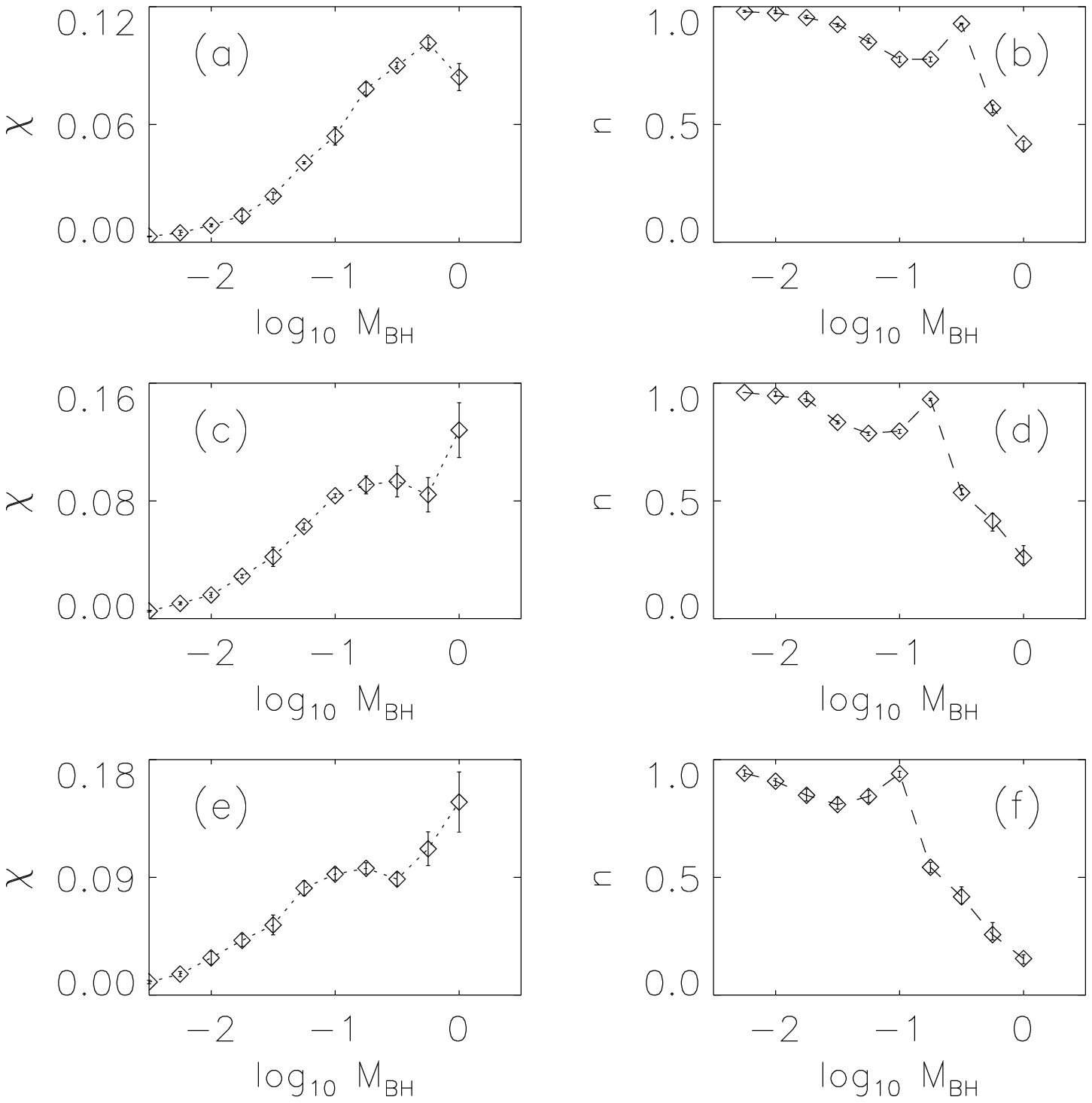}
           }
        \begin{minipage}{10cm}
        \end{minipage}
        \vskip -0.3in\hskip -0.0in
\caption{ 
(a) Estimates of the largest Lyapunov exponent for chaotic orbits evolved in
the potential with $a^2=1.0$, $b^2=1.25$, $c^2=0.75$, ${\epsilon}=10^{-2}$, and
$E=1.0$ as a function of $M_{BH}$. (b) The fraction of chaotic orbits from a 
representative ensemble with the same $a$, $b$, $c$, ${\epsilon}$, and $E$ as
a function of $M_{BH}$. (c) and (d) The same as (a) and (b) but for $E=0.6$.
(e) and (f) The same as (a) and (b) for $E=0.4$.}
\vspace{0.0cm}
\label{FIG5}
\end{figure}

This interpretation is corroborated by the fact that similar distributions
of short time Lyapunov exponents can be generated from a single initial 
condition integrated for very long times. Specifically, if a single initial 
condition with the same energy be integrated for a time $k \times 256$ and
short time exponents computed separately for each $t=256$ segment, the
resulting $N[{\chi}]$ is often nearly indistinguishable from the $N[{\chi}]$ 
generated from an ensemble of initial conditions. 
This is, {\em e.g.}, evident from Figure~\ref{FIG4},
which exhibits distributions generated for two different initial conditions
with $E=1.0$ integrated with $a^{2}=1.25$, $b^{2}=1.0$, $c^{2}=0.75$, 
${\epsilon}=10^{-2}$, and $M_{BH}=10^{-3}$. Each initial condition was 
integrated for a total time $t=8000 \times 256$ with energy conserved to 
better than one part in $10^{6}$. One corresponded at early times to a wildly
chaotic orbit with ${\chi}(t=256){\;}{\approx}{\;}0.06$; the other 
corresponded initially to a very `sticky,' near-regular orbit.

Figure~\ref{FIG5} 
exhibits estimates of ${\chi}$, the largest Lyapunov exponent, and 
$n(M_{BH})$, the relative fraction of obviously chaotic orbits, as functions 
of $M_{BH}$
for representative ensembles computed with $a^{2}=1.25$, $b^{2}=1.0$, 
$c^{2}=0.75$, and ${\epsilon}=10^{-2}$ for three different energies, namely 
$E=1.0$, $E=0.6$, and $E=0.4$. Only the data for $M_{BH}{\;}{\ge}{\;}10^{-2.5}$
are exhibited since, as noted above, for smaller values it becomes extremely
difficult to determine whether any given orbit is regular or chaotic. However,
it {\em does} seem clear that there is little or no chaos when the black hole 
mass is as small as $M_{BH}=10^{-4}$, and that there is a comparatively 
rapid increase in the importance of chaos for $M_{BH}{\;}{\sim}{\;}10^{-3}$.
This rapid increase is probably not generic. Rather, as discussed in Section 
4, the fact that chaos appears to `turn on' abruptly likely reflects the 
fact that, for this potential, all the unperturbed ($M_{BH}=0$) orbits with
fixed energy $E$ have exactly the same natural frequencies and densely sample
the same region in configuration space, so that `if you've
seen one orbit you've seen them all.' 

If the black hole mass be increased from $M_{BH}=10^{-2.5}$ to larger values,
the overall abundance of chaos tends to decrease. However, this decrease is 
{\em not} completely uniform. For all three energies, one observes a 
pronounced spike in the distribution $n(M_{BH})$ where the relative measure of 
chaotic orbits increases abruptly. The existence of these spikes indicates 
that the relative abundance of chaos does not exhibit a completely trivial 
dependence on $M_{BH}$. The fact that the locations of these spikes varies 
smoothly with energy suggests strongly that this nontrivial variation is 
a resonance phenomenon.

Similarly, for fixed $a$, $b$, $c$, ${\epsilon}$, and $E$, the largest Lyapunov
exponent, as defined in an asymptotic $t\to\infty$ limit, does not exhibit a 
completely trivial dependence on $M_{BH}$. In particular, ${\chi}$ is not a 
monotonically increasing function of $M_{BH}$. However, for ensembles 
integrated for times $t{\;}{\sim}{\;}100-800t_{D}$, it appears that the 
largest short time exponents observed for any of the orbits {\em do} increase
monotonically with increasing $M_{BH}$. The fact that, for certain ranges of 
$M_{BH}$, the asymptotic ${\chi}$ decreases reflects the fact that, for these 
values of $M_{BH}$, it is especially likely for a chaotic orbit to be `nearly 
regular' for long times, during which the short time ${\chi}$ is comparatively 
small. That the locations of these `dips' varies smoothly with energy again 
suggests that some resonance phenomenon has come into play.

\begin{figure}
\centerline{
        \epsfxsize=8cm
        \epsffile{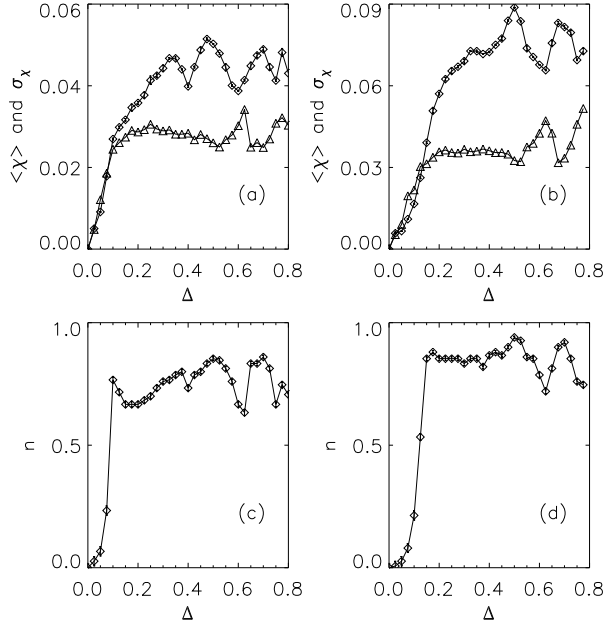}
           }
        \begin{minipage}{10cm}
        \end{minipage}
        \vskip -0.0in\hskip -0.0in
\caption{ 
(a) The average short time Lyapunov exponents, ${\langle}{\chi}{\rangle}$
(diamonds), and the associated dispersions, ${\sigma}_{\chi}$ (triangles), for 
representative ensembles with ${\epsilon}=10^{-2}$, $E=1.0$, $M_{BH}=0.1$, and 
$a^2:b^2:c^2=1+{\Delta}:1:1-{\Delta}$ for variable ${\Delta}$. (b) The same
for an ensemble with $E=0.6$. (c) The fraction $n$ of wildly chaotic orbits for
the $E=1.0$ ensemble in (a). (d) The same for the $E=0.6$ ensemble.}
\label{FIG6}
\end{figure}

Overall, chaos seems to become more pronounced for larger deviations from
axisymmetry but, once again, the trend is not completely uniform. Rather, it
is again possible to observe dips and/or spikes in the relative measure of
chaotic orbits. This is, {\em e.g.}, illustrated by Figure~\ref{FIG6}, which 
exhibits 
${\langle}{\chi}{\rangle}$, the mean value of the largest Lyapunov exponent,
${\sigma}_{\chi}$, the associated dispersion, and $n$, the relative abundance
of strongly chaotic orbits for sequences of models with fixed $E$, 
$M_{BH}=0.1$, ${\epsilon}=10^{-2}$, and 
$a^{2}:b^{2}:c^{2}=1+{\Delta}:1:1-{\Delta}$ 
with $0{\;}{\le}{\;}{\Delta}{\;}{\le}{\;}0.8$. The two left panels were 
computed for
ensembles with $E=1.0$; the two right panels were computed for $E=0.6$.

\section{Spectral Properties of Orbits and Orbit Ensembles}
\subsection{WHAT WAS COMPUTED}
The experiments described here aimed to provide additional insights into the
origins of chaos in the potential (1) by examining the Fourier spectra of 
representative orbits and orbit ensembles; testing the validity of a 
perturbative analysis which views 
$V_{p}=-M_{BH}/\sqrt{{r^{2}+{\epsilon}^{2}}}$ as a perturbation of the 
unperturbed oscillator potential; and examining how, for individual orbits, 
the degree of instability exhibited at different points correlates with the
location of the orbit in configuration space.

Two classes of experiments were considered. The first involved computing 
Fourier spectra for each orbit in several of the ensembles described in the
preceding section. This facilitated an improved understanding of how various
orbits with the same $E$, $a$, $b$, $c$, ${\epsilon}$, and $M_{BH}$ can differ 
one from another.

The other class of experiments explored the question of how, for a fixed set
of initial conditions, evolved with the same choices of $a$, $b$, $c$, and 
${\epsilon}$, varying the central mass $M_{BH}$ impacts bulk statistical 
properties. This entailed selecting fifty representative phase space points, 
uniformly sampling the interval $0.16<r<1.4$, with $v_{y}=v_{z}=0$ and $v_{x}$ 
so chosen that $E=1.0$ for $M_{BH}=0.1$; and then evolving these into the 
future for a time $t=2048$ with $a^{2}=1.25$, $b^{2}=1.0$, $c^{2}=0.75$, and 
${\epsilon}=10^{-2}$, varying the black hole mass 
incrementally from $M_{BH}=0$ to $M_{BH}=0.4$ in steps of $0.01$. An analysis
of the resulting Fourier spectra facilitated a quantitative characterisation
of the extent to which perturbation theory correctly predicts changes in 
frequencies. A comparison of short time Lyapunov exponents and properties 
of the Fourier spectra for the orbits corroborated the expectation that, as 
for other two- and three-dimensional potentials 
({\em cf.} \citeauthor{KEB} \citeyear{KEB}, \citeauthor{SEK} \citeyear{SEK}),
orbit segments with 
larger Lyapunov exponents typically have more `complex' Fourier spectra, with 
substantial power at a larger number of different frequencies.

The notion of `complexity' exploited here differs slightly from that adopted
in earlier papers. In those papers, the complexity $n(k)$ was defined as the
number of frequencies required in a discrete Fourier spectrum to capture a
fixed fraction $k$ of the total power. Here the complexity was defined
instead as equaling the number of frequencies which contained power greater
than or equal to a fixed fraction $k$ of the power at the peak frequency. 
The results exhibited here involved the choice $k=0.7$ and a complexity
\begin{equation}
n({\omega})={1\over 3}\,{\bigl[}
n_{x}({\omega})+n_{y}({\omega})+n_{x}({\omega}){\bigr]}
\end{equation}
where, {\em e.g.}, $n_{x}$ denotes the number of frequencies for motion in the 
$x$-direction. The alternative prescription exploited here reflects the fact
that there is no obvious unique definition of complexity, and that different
definitions yield comparable results. 

The analytic computation of modified frequencies was effected using standard
second order perturbation theory ({\em cf.} \citeauthor{lal} \citeyear{lal}).
Relevant formulae are summarised in an Appendix.
\subsection{WHAT WAS FOUND}
\begin{figure}
\centerline{
        \epsfxsize=8cm
        \epsffile{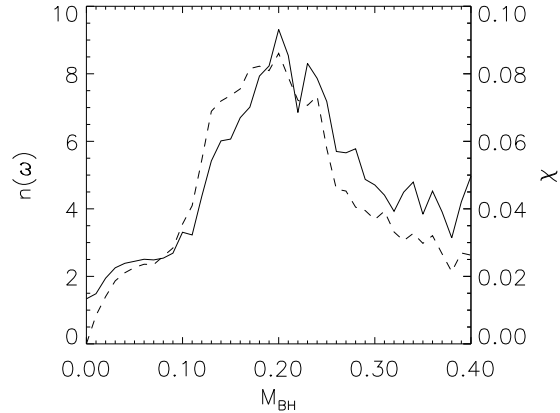}
           }
        \begin{minipage}{10cm}
        \end{minipage}
        \vskip -0.4in\hskip -0.0in
\caption{ 
$n({\omega})$, the mean number of frequencies with amplitude greater than 
$0.7$ times the frequency with peak power (solid curve), and ${\chi}$, the 
mean short time Lyapunov exponent (dashed curve), computed for the same set of
fifty different initial conditions with  $E=1.0$ evolved in the potential
(1) for $t=2048$ with $a^{2}=1.25$, $b^{2}=1.0$, and $c^{2}=0.75$,
for a range of black hole masses $0{\;}{\le}{\;}M_{BH}{\;}{\le}{\;}0.4$. 
}
\label{FIG7}
\end{figure}

Overall, the complexities of different orbit segments correlate well with the 
values of the largest short time Lyapunov exponent ${\chi}$: orbits for which 
power is spread over a larger number of frequencies tend systematically to 
have larger values of ${\chi}$. That this should be the case for different 
orbits evolved in the same potential with the same energy is hardly surprising,
given results for other three-dimensional potentials 
({\em cf.} \citeauthor{SEK} \citeyear{SEK}).
Less obvious, however, but also true is that such 
a correlation persists even when comparing ensembles with different values
of $M_{BH}$. This is, {\em e.g.}, evident from Figure~\ref{FIG7}, which plots 
as a function
of $M_{BH}$ the average values of $n({\omega})$ (solid curve) and ${\chi}$ 
(dashed curve) for the aforementioned fifty orbit ensembles. It is clear that 
the two plots are quite similar, each peaking at a mass $M_{BH}{\;}{\sim}{\;}
0.2$ where, consistent with Figures 2 and 3, the relative measure of wildly 
chaotic 
orbits is especially high.

\begin{figure}
\centerline{
        \epsfxsize=8cm
        \epsffile{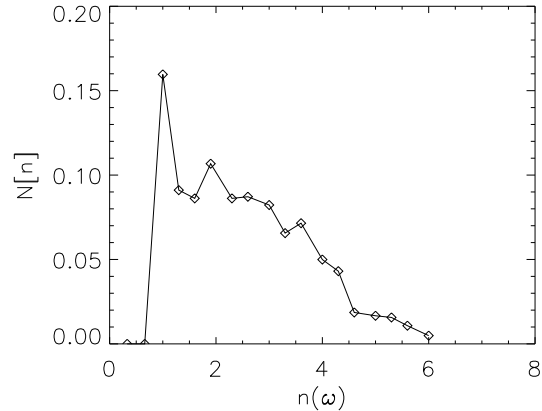}
           }
        \begin{minipage}{10cm}
        \end{minipage}
        \vskip -0.3in\hskip -0.0in
\caption{ 
The relative number of orbits with different complexities $n(\omega)$, 
computed for the same ensemble used to generate Figure~\ref{FIG3} b.}
\label{FIG8}
\end{figure}

For most values of $M_{BH}$ there are relatively few orbits with very large 
values of ${\chi}$: most of the orbits are either regular or nearly regular.
For example, for $E=1.0$, $a^{2}=1.25$, $b^{2}=1.0$, and $c^{2}=0.75$, there
are large measures of wildly chaotic orbits only for $M_{BH}$ between about 
$10^{-1.5}$ and $10^{-0.5}$. For much larger or smaller values $M_{BH}$ one 
finds that $N[{\chi}]$, the distribution of short time Lyapunov exponents, is a
decreasing function of ${\chi}$. Thus, it is hardly surprising that $N[n]$,
the distribution of complexities, should be a decreasing function of $n$. 
Less obvious, however, but also true is that even for a collection of wildly
chaotic orbits, where $N[{\chi}]$ peaks at some intermediate value of ${\chi}$,
$N[n]$ can be a sharply decreasing function of $n$. This is, {\em e.g.}, 
evident from Figure~\ref{FIG8}, 
which plots $N[n]$ for the same orbit ensemble used to 
generate the distribution $N[{\chi}]$ in Figure~\ref{FIG3} b. In this plot, the
orbits with $n({\omega})<2$ are almost all regular or nearly regular, whereas 
the orbits with $n({\omega})>2$ are almost all wildly chaotic. The obvious 
point is that, for $n({\omega})>2$, the distribution $N[n]$ decreases in a 
near-uniform fashion, even though the corresponding distribution $N[{\chi}]$ 
has a pronounced peak at ${\chi}{\;}{\sim}{\;}0.06$.

Second order perturbation theory can be expected to work reasonably well only 
when $M_{BH}$ is relatively small; and, for this reason, comparisons between
analytic predictions and numerical computations were only effected for 
$M_{BH}=0.4$ and less. Even for such small values, however, subtleties 
remain. By construction, perturbation theory computes shifts in three basic
frequencies which, for $M_{BH}=0$, reduce to ${\omega}=a$, $b$, and $c$.
However, one discovers oftentimes that, even for regular orbits, $M_{BH}>0$
implies spectra with power at more than three frequencies; and, for fully
chaotic orbits, the spectra should, strictly speaking, be continuous 
({\em cf.} \citeauthor{tab} \citeyear{tab}),
although most of the power may be concentrated at or near a 
small number of frequencies. For this reason, identifying the relative error 
between the analytics and numerical computations requires a concrete 
prescription to identify what one really means by the `principal' frequencies 
for oscillations in the $x$, $y$, and $z$ directions. As a practical matter, 
a principal frequency like ${\omega}_{x}$ was extracted by identifying those 
frequencies in the Fourier spectrum which have at least 70\% as much power as 
that contained in the frequency with maximum power, and then computing 
a power-weighted 
average value for these frequencies. A net error between 
analytics and numerics for a 
single orbit was derived by averaging over the error associated with the three 
different frequencies, ${\omega}_{x}$, ${\omega}_{y}$, and ${\omega}_{z}$.

\begin{figure}
\centerline{
        \epsfxsize=8cm
        \epsffile{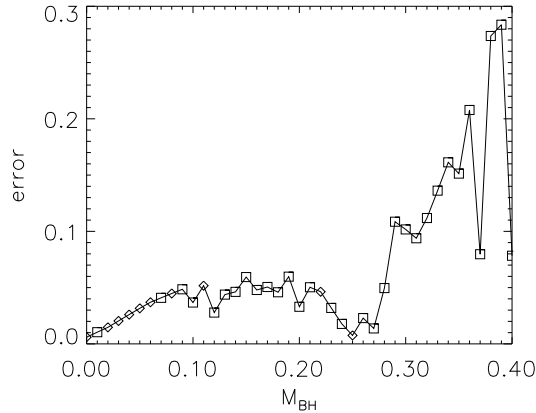}
           }
        \begin{minipage}{10cm}
        \end{minipage}
        \vskip -0.3in\hskip -0.0in
\caption{ 
The fractional error between the values of the `principal' frequency 
${\omega}_{x}$ computed numerically and calculated analytically for a single
orbit with $E=1.0$ evolved in the potential (1) for $t=2048$ with 
$a^{2}=1.25$, $b^{2}=1.0$, and $c^{2}=0.75$ for a range of black hole masses 
$0{\;}{\le}{\;}M_{BH}{\;}{\le}{\;}0.4$. Squares denote chaotic orbits, whereas
diamonds denote regular or nearly regular orbits.
}
\label{FIG9}
\end{figure}

Overall, it was found that, for $M_{BH}<0.2$ or so, the typical discrepancy 
between the analytics and numerics is quite small, less than or of order 5\%. 
However, as illustrated in Figure~\ref{FIG9}, which exhibits the error for 
${\omega}_{x}$ for a single
representative orbit, the typical errors become substantially larger for 
$M_{BH}>0.2$ or so. One might suppose that analytic predictions of the
`principal frequencies' for chaotic orbits would tend to be much worse than 
for regular orbits, and that the abrupt increase in relative error for 
$M_{BH}>0.25$ reflects the onset of chaos. This, however, is not the whole
story. For this initial condition, almost all the orbits with $M_{BH}>0.08$ 
are obviously chaotic, but the fractional errors are all less than $0.06$ for 
$M_{BH}<0.3$.

\begin{figure}
\centerline{
        \epsfxsize=8cm
        \epsffile{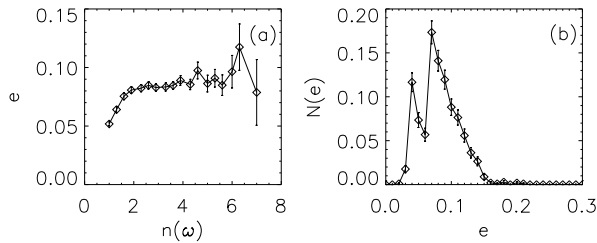}
           }
        \begin{minipage}{10cm}
        \end{minipage}
        \vskip -2.3in\hskip -0.0in
\caption{ 
(a) The average fractional error as a function of $n({\omega})$, the number of 
frequencies with amplitude greater than $0.7$ times the frequency with peak 
power for the orbit ensemble used to generate Figure~\ref{FIG7}. The error 
bars reflect
the spread of values observed for orbits with the same $n({\omega})$.
(b) The relative fraction of the orbits in the ensemble with different 
fractional errors.}
\label{FIG10}
\end{figure}

Nevertheless, it is apparent that, for fixed value of $M_{BH}$, the typical 
error grows systematically with increasing complexity, which is hardly 
surprising in light of the fact that, for large complexity, the `principal' 
frequency really involves an average over several different frequencies. This 
is, {\em e.g.}, evident in the first panel of Figure~\ref{FIG10}, which 
exhibits the average error as a function of $n({\omega})$ for the same orbit 
ensemble used to generate Figure~\ref{FIG7}. That perturbation theory is more 
successful overall in accounting for regular and nearly regular orbit segments 
than for wildly chaotic segments is evident from the second panel in 
Figure~\ref{FIG10}, which plots 
$N[e]$, the distribution of relative errors. That the distribution is sharply 
bimodal is consistent with the fact that, as regards the success of 
perturbation theory, the orbits divide into two distinct populations, one, 
comprised mostly of regular orbits, for which perturbation theory is quite 
reliable, and another, comprised mostly of wildly chaotic orbits, for which 
perturbation theory works somewhat less well.

\begin{figure}
\centerline{
        \epsfxsize=8cm
        \epsffile{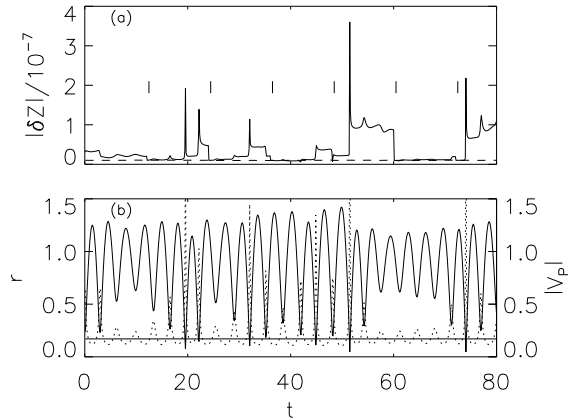}
           }
        \begin{minipage}{10cm}
        \end{minipage}
        \vskip -0.4in\hskip -0.0in
\caption{ 
(a) The solid curve shows $|{\delta}Z|(t)$, the magnitude of the phase space 
distance between a perturbed and unperturbed chaotic orbit evolved in the 
potential (1) with $a^{2}=1.25$, $b^{2}=1.0$, $c^{2}=0.75$, $M_{BH}=0.15$, 
and $E=0.75$. For comparison, the dashed line shows $|{\delta}Z|$ for a 
regular 
orbit. The orbit was renormalised at intervals ${\delta}t=12.0$ at points
indicated by vertical stripes in the plot. (b) The solid curve 
shows $r(t)$, the distance from the origin, for the same orbit at the same 
times. The dotted line shows the magnitude of the perturbation, 
$|V_{p}|=M_{BH}/{\sqrt{ r^{2}+{\epsilon}^2}}$. The vertical line corresponds to
a value $r=0.17$, for which $|V_{p}|{\;}{\approx}{\;}0.88$.}
\label{FIG11}
\end{figure}

But precisely {\em why} does the observed chaos actually arise? A detailed
examination of the local stability of individual orbits provides a simple and
compelling answer. When an orbit is far from the center, the central mass 
$M_{BH}$ has only a comparatively minimal effect and the orbit is nearly 
regular. However, when the orbit comes too close to the center and $M_{BH}$ 
contributes significantly to the total potential experienced by the orbit, 
the combination of the incompatible symmetries leads to a resonance overlap 
and a significant exponential instability. This is, {\em e.g.}, illustrated 
in Figure~\ref{FIG11}, which exhibits a representative piece of a wildly 
chaotic orbit 
segment with $E=0.75$ evolved in the potential (1) with $a^{2}=1.25$, 
$b^{2}=1.0$, $c^{2}=0.75$, and $M_{BH}=0.15$. Here the solid curve in
the top panel displays the phase space separation $|{\delta}Z(t)|$ between the
original orbit and a perturbed orbit displaced originally by a distance 
$|{\delta}Z|=10^{-8}$ and renormalised in the usual way 
({\em cf.} \citeauthor{lal} \citeyear{lal})
at intervals ${\delta}t=12.0$. The dashed curve shows an
analogous plot of $|{\delta}Z|$ for a regular orbit. The bottom panel exhibits
$r(t)$, the distance from the origin (solid curve), and 
$|V_{p}|=M_{BH}/\sqrt{{r^{2}+{\epsilon}^{2}}}$, the magnitude of the central 
mass perturbation. The obvious point is that most of the time the perturbed
and unperturbed orbit remain very close together, with little if any 
systematic exponential divergence; but that when $r<0.17$ or so, so that
$|V_{p}|$ becomes as large as ${\sim}{\;}0.88$, the two orbits tend to diverge
significantly. 

In this context, it should be noted that, even for the orbits which have been
characterised as `wildly chaotic,' the computed Lyapunov exponents are 
comparatively small when expressed in units of $t_{D}$. For `generic' chaotic
potentials, one discovers typically that the largest Lyapunov exponent 
typically assumes a value ${\sim}{\;}0.5t_{D}^{-1}-1.0t_{D}^{-1}$ which, for
$t_{D}{\;}{\sim}{\;}2-3$, implies values of ${\chi}$ in excess of $0.2$ or so.
However, only for very large central masses, $M_{BH}>0.3$ or so, did any of
the computed orbit segments have values of ${\chi}$ as large as $0.2$. Even 
the so-called wildly chaotic orbits tend to be comparatively regular most of 
the time, the observed chaos resulting from those comparatively rare intervals 
when the orbits come comparatively close to the central mass. 
\section{Discussion}

The stickiness manifested by chaotic orbits in the potential explored in
this paper is strikingly reminiscent of what has been observed for the triaxial
analogues of the Dehnen potentials 
({\em cf.} \citeauthor{MaV1} \citeyear{MaV1}, \citeauthor{SaK} \citeyear{SaK}),
especially for those orbits which, in the absence of the
cusp and/or black hole, would correspond to box orbits. Introducing a cusp 
and/or a supermassive black hole into an
otherwise smooth triaxial potential can generate a considerable amount of
chaos, but individual orbit segments can be very nearly regular for very long
times. Stickiness appears to be a more important phenomenon in cuspy triaxial
potentials, both with and without a central black hole, than in many other 
chaotic three-dimensional potentials. 

The numerical experiments described in Section 2 -- especially the fact that
quantities like the relative measure of chaotic, as opposed to regular, orbits
can manifest a nontrivial dependence on both $M_{BH}$ and axis ratio -- 
suggests strongly that the chaos associated with this potential results from a 
resonance overlap between the natural frequencies of the unperturbed oscillator
and the natural frequencies of the central black hole. An examination of the 
local stability of individual orbits indicates that, for intermediate black 
hole masses, the orbits are only very unstable when they are sufficiently 
close to the black hole that $|V_{o}|$ and $|V_{p}|$ become comparable in 
magnitude. The computations for this simple potential thus corroborate the
conclusions of
\citeauthor{VaM} \shortcite{VaM}
based on a detailed investigation of 
the spectral properties of orbits in the triaxial Dehnen potentials. However,
it is also evident that the chaos arises in response to a combination of
incompatible symmetries: the Plummer and oscillator potentials are each
integrable separately but, when combined with comparable magnitudes, can yield
strongly chaotic behaviour.

The simplicity of the potential (1) was important because it made possible
the analytic calculations described in Section 3. However, this potential 
{\em is} artificial in the sense that, for fixed energy $E$, every orbit 
unperturbed by a central black hole has the same natural frequencies, a fact
that renders the conclusions of the paper anomalous in two significant
respects: (1) As is clear, {\em e.g.}, from Figures~\ref{FIG2} and ~\ref{FIG3},
as $M_{BH}$ 
increases there is an extremely abrupt transition from a phase space 
hypersurface with essentially  no chaos to a phase space that is almost 
completely chaotic. If the harmonic oscillator component of the potential is 
replaced by an anharmonic oscillator, {\em e.g.}, by allowing for quartic 
corrections, the natural frequencies of different unperturbed orbits with the 
same energy are no longer identical, and one would anticipate the transition 
from little chaos to much chaos to become more gradual. (2) If the oscillator 
is made anharmonic, the nontrivial structures exhibited by $n(M_{BH})$ and
$n({\Delta})$ might also be expected to disappear. Because an ensemble of 
fixed 
energy now involves unperturbed orbits with different frequencies, a plot of 
$n(M_{BH})$ can be understood as involving a nontrivial superposition of 
orbits with different unperturbed frequencies. 

\begin{figure}
\centerline{
        \epsfxsize=8cm
        \epsffile{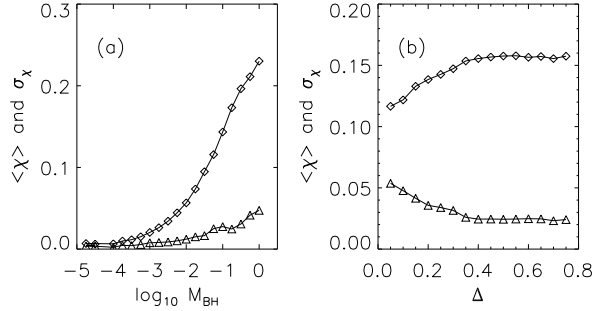}
           }
        \begin{minipage}{10cm}
        \end{minipage}
        \vskip -1.8in\hskip -0.0in
\caption{ 
(a) The average short time Lyapunov exponents, ${\langle}{\chi}{\rangle}$
(diamonds), and the associated dispersions, ${\sigma}_{\chi}$ (triangles), for 
representative ensembles with ${\epsilon}=10^{-2}$, $E=1.0$, and 
$a^2:b^2:c^2=1.25:1.00:0.75$ evolved in the anisotropic potential (3) with
${\alpha}=1.0$ and variable $M_{BH}$. (b) The average short time Lyapunov 
exponents, ${\langle}{\chi}{\rangle}$ (diamonds), and the associated 
dispersions, ${\sigma}_{\chi}$ (triangles), for 
representative ensembles with ${\epsilon}=10^{-2}$, $E=1.0$, $M_{BH}=0.1$, and 
$a^2:b^2:c^2=1+{\Delta}:1:1-{\Delta}$ for variable ${\Delta}$.}
\label{FIG12}
\end{figure}

These expectations were confirmed by repeating the computations described in
this paper for orbits in the generalised potential
\begin{displaymath}
V(x,y,z)={1\over 2}(a^{2}x^{2}+b^{2}y^{2}+c^{2}z^{2})
\;\;\;\;\;\;\;\;\;\;\;\;\;\;\;\;\;\;
\end{displaymath}
\begin{equation}
\;\;\;\;\;\;\;\;\;\;\;\;\;\;\;\;\;\;
+{{\alpha}\over 4}(a^{2}x^{4}+b^{2}y^{4}+c^{2}z^{4})-
{M_{BH}\over \sqrt{r^{2}+{\epsilon}^{2}}},
\end{equation}
for a variety of different values of ${\alpha}$. Data for the representative
value ${\alpha}=1$ are exhibited in Figure~\ref{FIG12}. Here the first panel 
exhibits
the mean ${\langle}{\chi}{\rangle}$ and dispersion ${\sigma}_{\chi}$ for 
$2000$ orbit ensembles with $E=1.0$, $a^{2}=1.25$, $b^{2}=1.0$, $c^{2}=0.75$
and ${\epsilon}^{2}=10^{-4}$ for different values of $M_{BH}$. The second
panel exhibits the same quantities for ensembles with $E=1.0$, $M_{BH}=0.1$,
${\epsilon}^{2}=10^{-4}$, and $a^{2}:b^{2}:c^{2}=1+{\Delta}:1:1-{\Delta}$ 
for different values of ${\Delta}$. It is evident from panel (a) that that
the transition to chaos is more gradual than was observed in the harmonic
potential. Moreover, only for black hole masses somewhat smaller than 
$M_{BH}=10^{-4}$ do all the orbits seem completely regular. And similarly, it 
is clear that ${\langle}{\chi}{\rangle}$ and ${\sigma}_{\chi}$ vary smoothly 
with $M_{BH}$ and ${\Delta}$, the resonant behaviour observed for ${\alpha}=0$
having washed out because the ensembles now involve unperturbed orbits with
a variety of frequencies.

\appendix
The Hamiltonian appropriate for motion in the potential (1) can be written
in the form $H=H_{0}+H_{1}$, where
\begin{equation}
H_{0}={1\over 2}(P_{x}^{2}+P_{y}^{2}+P_{z}^{2})+
{1\over 2}(A_{x}^{2}Q_{x}^{2}+A_{y}^{2}Q_{y}^{2}+A_{z}^{2}Q_{z}^{2})
\end{equation}
is integrable and
\begin{equation}
H_{1}=-{M_{BH}\over \sqrt{{Q_{x}^{2}+Q_{y}^{2}+Q_{z}^{2}+{\epsilon}^{2}}}}
\end{equation}
may be viewed as a perturbation that breaks integrability. Because the 
integrable component $H_{0}$ is separable, it is easy to compute
the generating functional $W(Q_{1},J_{1},Q_{2},J_{2},Q_{3},J_{3})$, which 
can be written as
\begin{equation}
W=\sum_{i=1}^{3}\,{1\over 2}Q_{i}\sqrt{2J_{i}-A_{i}^{2}Q_{i}^{2}}+
{J_{i}\over A_{i}}\,\sin^{-1}{\Bigl(}{A_{i}x_{i}\over \sqrt{2J_{i}}}{\Bigr)}
\end{equation}
where $J_{i}$ denote the actions and the index $i$ ranges over $x$, $y$, and 
$z$. Given $W$, it is easy to relate the original canonical variables to the
actions $J_{i}$ and a corresponding set of angles ${\theta}_{i}$, in terms of 
which the Hamiltonian reduces to $H=H_{0}+H_{1}$, where
\begin{equation}
H_{0}=J_{1}+J_{2}+J_{3}
\end{equation}
and
$$H_{1}=-M_{BH}\times \;\;\;\;\;\;\;\;\;\;\;\;\;\;\;\;\;\;\;\;\;\;\;\;\;\;
\;\;\;\;\;\;\;\;\;\;\;\;\;\;\;\;\;\;\;\;\;\;\;\;\;\;\;\;\;\;\;\;\;\;\;
\;\;\;\;\;\;\;\;$$
\begin{equation}
\,{\biggl[}
{2J_{1}\over A_{1}^{2}}\sin^{2}A_{1}{\theta}_{1} +
{2J_{2}\over A_{2}^{2}}\sin^{2}A_{2}{\theta}_{2} +
{2J_{3}\over A_{3}^{2}}\sin^{2}A_{3}{\theta}_{3}+{\epsilon}^{2}{\biggr]}^{-1/2}
.\end{equation}
At this stage, one needs to perform a triple integral, averaging each of
the three angles over the range $0<{\theta}<2{\pi}$. Unfortunately, however,
this is difficult to do in closed form. Nevertheless, one can proceed 
perturbatively by expanding $H_{1}$ in a Taylor series around some point
$({\theta}_{1},{\theta}_{2},{\theta}_{3})=(q_{1},q_{2},a_{3})$. The result of 
such a computation is that the perturbed frequencies ${\omega}_{i}$ satisfy
\begin{eqnarray}
{{\omega}_{i}\over A_{i}} & = &1 +
{{\partial}\over {\partial}J_{i}}{\Biggl[}
H_{1}(M_{BH},q_{1},q_{2},q_{3},J_{1},J_{2},J_{3})+ \nonumber\\
& + &\sum_{j=1}^{3}\,{\biggl(} {{\partial}H_{1}\over {\partial}q_{j}} {\biggr)}
\,({\pi}-q_{j})
\nonumber\\
& + & {1\over 2}\sum_{j=1}^{3}\,{\biggl(} 
{{\partial}^{2}H_{1}\over {\partial}q_{j}^{2}}{\biggr)}
{\Bigl(} {4\over 3}{\pi}^{2}-2{\pi}q_{j} + q_{j}^{2} {\Bigr)} \nonumber\\
& + & \sum_{j<k=1}^{3}\, {\biggl(}
{{\partial}^{2}H_{1}\over {\partial}q_{j}{\partial}q_{k}} {\biggr)}
{\bigl[} ({\pi}-q_{j})({\pi}-q_{k}) {\bigr]}{\Biggr]} .
\end{eqnarray}
It is natural to choose the $q_{i}$'s at or near values for which 
$A_{i}q{i}={\pm}{\pi}/2$, for which values the magnitude of $H_{1}$ is 
minimised.

\acknowledgements
The authors were supported in part by the National Science Foundation grant
AST-0070809 and by a grant from the Institute for Geophysics and Planetary 
Physics at Los Alamos National Laboratory. HEK also acknowledges useful 
discussions with Chris Hunter and Christos Siopis.

\end{article}
\end{document}